\documentclass{sig-alternate}

\DeclareMathAlphabet{\mathitbf}{OML}{cmm}{b}{it}

\usepackage{underscore}

\begin{document}

\title{Building a Balanced k-d Tree in O(kn log n) Time} 


\author{Russell A. Brown}

\date{01 April 2015}
\maketitle
\begin{abstract}

The original description of the \emph{k}-d tree recognized that rebalancing techniques, such as are used to build an AVL tree or a red-black tree, are not applicable to a \emph{k}-d tree.  Hence, in order to build a balanced \emph{k}-d tree, it is necessary to find the median of the data for each recursive subdivision of those data.  The sort or selection that is used to find the median for each subdivision strongly influences the computational complexity of building a \emph{k}-d tree.

This paper discusses an alternative algorithm that builds a balanced \emph{k}-d tree by presorting the data in each of $k$ dimensions prior to building the tree.  It then preserves the order of these $k$ sorts during tree construction and thereby avoids the requirement for any further sorting.  Moreover, this algorithm is amenable to parallel execution via multiple threads.  Compared to an algorithm that finds the median for each recursive subdivision, this presorting algorithm has equivalent performance for four dimensions and better performance for three or fewer dimensions.

\end{abstract}




\section{Introduction} 
\label{sec:introduction}

Bentley introduced the \emph{k}-d tree as a binary tree that stores \emph{k}-dimensional data \cite{Bentley}.  Like a standard binary tree, the \emph{k}-d tree subdivides data at each recursive level of the tree.  Unlike a standard binary tree that uses only one key for all levels of the tree, the \emph{k}-d tree uses $k$ keys and cycles through these keys for successive levels of the tree.  For example, to build a \emph{k}-d tree from three-dimensional points that comprise $\left(x,y,z\right)$ coordinates, the keys would be cycled as $x,y,z,x,y,z...$  for successive levels of the \emph{k}-d tree.  A more elaborate scheme for cycling the keys chooses the coordinate that has the widest dispersion or largest variance to be the key for a particular level of recursion \cite{Friedman}.

Due to the use of different keys at successive levels of the tree, it is not possible to employ rebalancing techniques, such as are used to build an AVL tree \cite{Adelson} or a red-black tree \cite{Bayer,Guibas}, when building a \emph{k}-d tree.  Hence, the typical approach to building a balanced \emph{k}-d tree finds the median of the data for each recursive subdivision of the data.  Bentley showed that if the median of $n$ elements were found in $O\left(n\right)$ time, it would be possible to build a depth-balanced \emph{k}-d tree in $O\left(n \log n\right)$ time.  Blum \emph{et al.} proposed an elegant but slightly complicated algorithm that finds the median in guaranteed $O\left(n\right)$ time \cite{Blum}.  Quicksort \cite{Hoare} finds the median in $O\left(n\right)$ time in the best case but in $O\left(n^2\right)$ time in the worst case \cite{Wirth}. Merge sort \cite{Neumann} and heap sort \cite{Williams} find the median in guaranteed $O\left(n \log n\right)$ time, which leads to $O\left(n \log^2 n\right)$ time for building a balanced \emph{k}-d tree \cite{Wald}.

An alternative approach to building a balanced \emph{k}-d tree presorts the data prior to building the tree and then avoids resorting for each recursive subdivision.  Two algorithms have been reported that sort triangles for three-dimensional graphics ray tracing and that have best-case complexity of $O\left(n \log n\right)$ but undetermined worst-case complexity \cite{Havran,Wald}.  The algorithm described in the present article presorts points in each of $k$ dimensions prior to building the \emph{k}-d tree, then maintains the order of the $k$ sorts when building a balanced \emph{k}-d tree, and thereby achieves a worst-case complexity of $O\left(kn \log n\right)$.  Procopiuc \emph{et al.} have outlined an algorithm \cite{Procopiuc} that appears to be similar to the algorithm described in the present article.

\section{Implementation}
\subsection{The $\mathitbf{O\left(kn \; \boldsymbol{\log} \; n\right)}$ Algorithm}
\label{sec:knlogn-algorithm}

\begin{figure}[h]
\centering
\centerline{\includegraphics*[trim = {0.09in, 0.20in, 0.07in, 0.00in}, clip, width=\columnwidth]{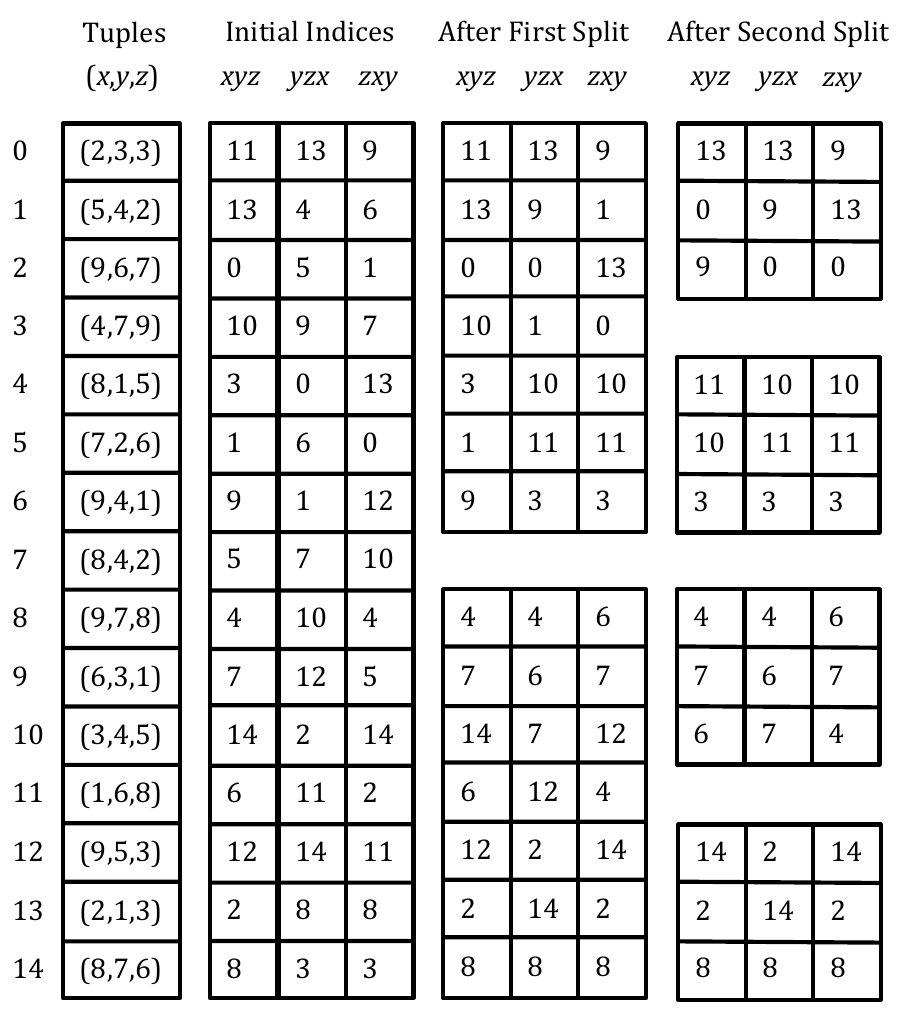}}
\caption{An $\mathitbf{\left(x,y,z\right)}$ tuple array and $\mathitbf{xyz}$-, $\mathitbf{yzx}$- and $\mathitbf{zxy}$-index arrays.}
\label{fig:IndexArrays}
\end{figure}

Consider the 15 $\left(x,y,z\right)$ tuples that are stored in elements 0 through 14 of the ``Tuples" array that is shown at the left side of Figure \ref{fig:IndexArrays}.  The \emph{k}-d tree-building algorithm begins by presorting the tuples in their $x$-, $y$- and $z$-coordinates via three executions of merge sort.  These three sorts do not in fact sort the $x$-, $y$- and $z$-coordinates by using these coordinates as sort keys entirely independently of one another; instead, $x$, $y$ and $z$ form the most significant portions of the respective super keys $x$:$y$:$z$, $y$:$z$:$x$ and $z$:$x$:$y$ that represent cyclic permutations of $x$, $y$ and $z$.  The symbols for these super keys use a colon to designate the concatenation of the individual $x$, $y$ and $z$ values.  Hence, for example, the symbol $z$:$x$:$y$ represents a super key wherein $z$ is the most significant portion of the super key, $x$ is the middle portion of the super key, and $y$ is the least significant portion of the super key.

The merge sorts employ super keys, instead of keys that are merely the individual $x$-, $y$- and $z$-coordinates, in order to detect and remove duplicate tuples, as will be explained later.  The merge sorts do not reorder the Tuples array; rather, they reorder three index arrays whose elements are indices into the Tuples array.  The initial order of these index arrays is established by the merge sorts and is shown in Figure \ref{fig:IndexArrays} in the $xyz$, $yzx$ and $zxy$ columns under ``Initial Indices".  In this figure, $xyz$, $yzx$ and $zxy$ are shorthand notations for the super keys $x$:$y$:$z$, $y$:$z$:$x$ and $z$:$x$:$y$ respectively. 

The $xyz$, $yzx$ and $zxy$ columns under ``Initial Indices" represent the initial order of the $xyz$-, $yzx$- and $zxy$-index arrays that indicate the results of the three merge sorts.  For example, elements 0, 1, ... 13, 14 of the $xyz$-index array contain the sequence 11, 13, ... 2, 8 that represents the respective tuples $\left(1,6,8\right)$; $\left(2,1,3\right)$; ... $\left(9,6,7\right)$; $\left(9,7,8\right)$ that were ordered via merge sort using the respective super keys 1:6:8, 2:1:3, ... 9:6:7, 9:7:8.  Similarly, elements 0, 1, ... 13, 14 of the $yzx$-index array contain the sequence 13, 4, ... 8, 3 that represents the respective tuples $\left(2,1,3\right)$; $\left(8,1,5\right)$; ... $\left(9,7,8\right)$; $\left(4,7,9\right)$ that were ordered via merge sort using the respective super keys 1:3:2, 1:5:8, ... 7:8:9, 7:9:4.  Lastly, elements 0, 1, ... 13, 14 of the $zxy$-index array contain the sequence 9, 6, ... 8, 3 that represents the respective tuples $\left(6,3,1\right)$; $\left(9,4,1\right)$; ... $\left(9,7,8\right)$; $\left(4,7,9\right)$ that were ordered via merge sort using the respective super keys 1:6:3, 1:9:4, ... 8:9:7, 9:4:7.

The next step of the \emph{k}-d tree-building algorithm partitions the $\left(x,y,z\right)$ tuples in $x$ using the $x$:$y$:$z$ super key that is specified by the median element of the $xyz$-index array under ``Initial Indices".  This median element is located at address 7 of this array; its value is 5 and specifies the tuple $\left(7,2,6\right)$ for which the $x$:$y$:$z$ super key is 7:2:6.  The partitioning does not reorder the Tuples array; instead, it reorders the $yzx$- and $zxy$-index arrays.  The $xyz$-index array requires no partitioning because it is already sorted in $x$.  However, the $yzx$- and $zxy$-index arrays require partitioning in $x$ using the $x$:$y$:$z$ super key 7:2:6 that is specified by the median element of the $xyz$-index array.

This partitioning is accomplished for the $yzx$-index array as follows.  The elements of the $yzx$-index array are retrieved in order of increasing address from 0 to 14.  The $x$:$y$:$z$ super key that is specified by each element of the $yzx$-index array is compared to the 7:2:6 super key that is specified by the median element of the $xyz$-index array.  Each element of the $yzx$-index array is copied to either the lower or upper half of a temporary index array, depending on the result of this comparison.  After all of the elements of the $yzx$-index array have been processed in this manner, the temporary index array replaces the $yzx$-index array and becomes the new $yzx$-index array that is depicted in Figure \ref{fig:IndexArrays} under ``After First Split."  The partitioning of the first six elements of the $yzx$-index array is discussed below and provides insight into the details of the \emph{k}-d tree-building algorithm.

The element at address 0 of the $yzx$-index array is 13 and specifies the tuple $\left(2,1,3\right)$ for which the $x$:$y$:$z$ super key is 2:1:3.  This super key is less than the median super key 7:2:6; hence, the element at address 0 of the $yzx$-index array is copied to address 0 in the new $yzx$-index array, which is the lowest address in the lower half of the new $yzx$-index array.  The element at address 1 of the $yzx$-index array is 4 and specifies the tuple $\left(8,1,5\right)$ for which the $x$:$y$:$z$ super key is 8:1:5.  This super key is greater than the median super key 7:2:6; hence, the element at address 1 of the $yzx$-index array is copied to address 8 in the upper half of the new $yzx$-index array, which is the lowest address in the upper half of the new $yzx$-index array.  The element at address 2 of the $yzx$-index array is 5 and specifies the tuple $\left(7,2,6\right)$ for which the $x$:$y$:$z$ super key is 7:2:6.  This super key equals the median super key 7:2:6; hence, the element at address 2 in the $yzx$-index array is ignored and not copied to the new $yzx$-index array.

The element at address 3 of the $yzx$-index array is 9 and specifies the tuple $\left(6,3,1\right)$ for which the $x$:$y$:$z$ super key is 6:3:1.  This super key is less than the median super key 7:2:6; hence, the element at address 3 of the $yzx$-index array is copied to address 1 in the lower half of the new $yzx$-index array, which is the second lowest address in the lower half of the new $yzx$-index array.  The element at address 4 of the $yzx$-index array is 0 and specifies the tuple $\left(2,3,3\right)$ for which the $x$:$y$:$z$ super key is 2:3:3.  This super key is less than the median super key 7:2:6; hence, the element at address 4 of the $yzx$-index array is copied to address 2 in the lower half of the new $yzx$-index array, which is the third lowest address in the lower half of the new $yzx$-index array.  The element at address 5 of the $yzx$-index array is 6 and specifies the tuple $\left(9,4,1\right)$ for which the $x$:$y$:$z$ super key is 9:4:1.  This super key is greater than the median super key 7:2:6; hence, the element at address 5 of the $yzx$-index array is copied to address 9 in the upper half of the new $yzx$-index array, which is the second lowest address in the upper half of the new $yzx$-index array.

Partitioning continues for the remaining eight elements of the $yzx$-index array in the manner that is described above.  The partitioning of the first six elements of the $yzx$-index array reveals that the $yzx$-index array has been partitioned in $x$ relative to the median element of the $xyz$-index array; this partitioning preserves the initial merge-sorted order in $y$:$z$:$x$ within the lower and upper halves of the new $yzx$-index array.

Next, the $zxy$-index array is partitioned in $x$ relative to the median element of the $xyz$-index array, which preserves the initial merge-sorted order in $z$:$x$:$y$ for the lower and upper halves of the new $zxy$-index array.  The reader is encouraged to audit the partitioning of the first few elements of the $zxy$-index array under ``Initial Indices" in order to verify that these elements are correctly assigned to the lower and upper halves of the new $zxy$-index array that is shown in Figure \ref{fig:IndexArrays} under ``After First Split."  Because the partitioning in $x$ preserves the initial merge-sorted order for the lower and upper halves of the $yzx$- and $zxy$-index arrays, there is no requirement for any further sorting after the $k$ initial merge sorts.

\begin{figure}[h]
\centering
\centerline{\includegraphics*[trim = {0.48in, 0.20in, 0.48in, 0.00in}, clip, width=\columnwidth]{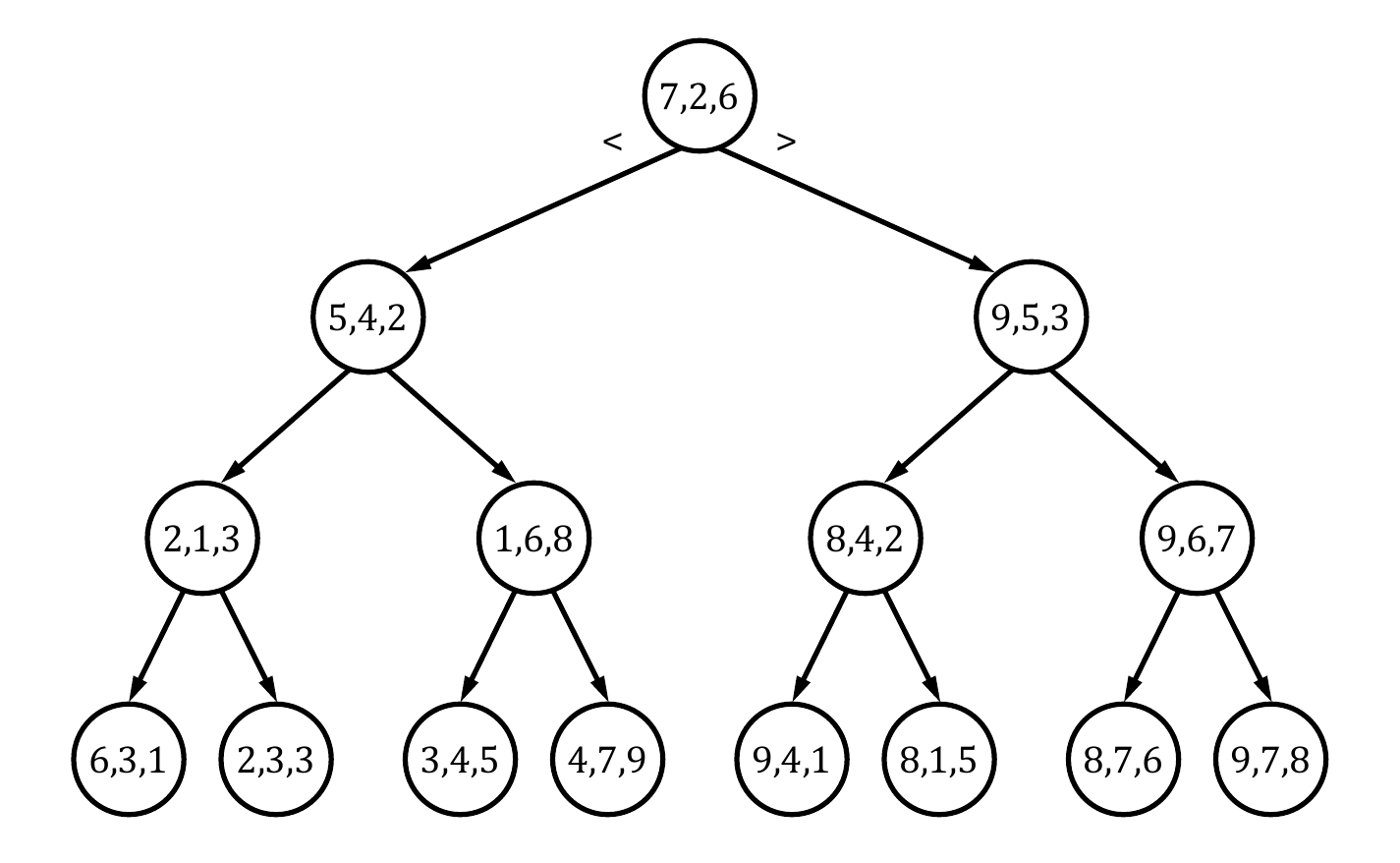}}
\caption{A \emph{k}-d tree that is built from the $\mathitbf{\left(x,y,z\right)}$ tuples of Figure  \ref{fig:IndexArrays}.}
\label{fig:FinalTree}
\end{figure}

The lower and upper halves of the new $xyz$-, $yzx$- and $zxy$-index arrays in Figure \ref{fig:IndexArrays} under ``After First Split" reveal that the index value 5 is absent from the lower and upper halves of these index arrays.  This value is absent from these index arrays because it is the value of the median element of the $xyz$-index array that specified the $x$:$y$:$z$ super key 7:2:6 relative to which the $yzx$- and $zxy$-index arrays were partitioned in $x$.  In order to record this partitioning, a reference to the tuple $\left(7,2,6\right)$ is stored in the root of the nascent \emph{k}-d tree, as shown in Figure \ref{fig:FinalTree}.

Next, the lower and upper halves of the $xyz$-, $yzx$- and $zxy$-index arrays are processed recursively and partitioned in $y$ to create the ``less than" and ``greater than" subtrees of the root of the \emph{k}-d tree.  Consider the lower half of the $yzx$-index array that is depicted in Figure \ref{fig:IndexArrays} under ``After First Split."  The median element of this array is located at address 3; its value is 1 and specifies the tuple $\left(5,4,2\right)$ for which the $y$:$z$:$x$ super key is 4:2:5.  The lower half of the $yzx$-index array is already sorted in $y$:$z$:$x$.  However, the lower halves of the $zxy$- and $xyz$-index arrays require partitioning in $y$ relative to the $y$:$z$:$x$ super key 4:2:5 that is specified by the median element of the lower half of the $yzx$-index array.  The reader is encouraged to verify the result of this partitioning by inspection of the first and second fourths of the new $xyz$-, $yzx$- and $zxy$-index arrays that are depicted in Figure \ref{fig:IndexArrays} under ``After Second Split."  The upper halves of the $zxy$- and $xyz$-index arrays are partitioned in a similar manner relative to the $y$:$z$:$x$ super key 5:3:9 that is formed from the tuple $\left(9,5,3\right)$ that is specified by the value 12 of the median element at address 11 of the upper half of the $yzx$-index array.  References to the tuples $\left(5,4,2\right)$ and $\left(9,5,3\right)$ are stored in the ``less than" and ``greater than" children of the root of the nascent \emph{k}-d tree, as shown in Figure \ref{fig:FinalTree}.

Recursion terminates when an index array comprises one, two or three elements.  In the case of one element, a reference to the corresponding tuple is stored in a new node of the \emph{k}-d tree.  For two or three elements, the elements are already in sorted order in the index array, so the determination of which tuple to reference from a new node of the \emph{k}-d tree and which tuple or tuples to reference from children of that node is trivial.  For example, consider the four fourths of the $zxy$-index arrays under ``After Second Split" in Figure \ref{fig:IndexArrays}.  Each fourth comprises three elements, so recursion terminates.  The tuples $\left(2,1,3\right)$ and $\left(1,6,8\right)$ that are specified respectively by the median elements 13 and 11 at addresses 1 and 5 of the $zxy$-index array are referenced by the respective ``less than" and ``greater than" children of node $\left(5,4,2\right)$ of the nascent \emph{k}-d tree.  Similarly, the tuples $\left(8,4,2\right)$ and $\left(9,6,7\right)$ that are specified respectively by the median elements 7 and 2 at addresses 9 and 13 of the $zxy$-index array are referenced by the respective ``less than" and ``greater than" children of node $\left(9,5,3\right)$ of the nascent \emph{k}-d tree.  The children and grandchildren of nodes $\left(5,4,2\right)$ and $\left(9,5,3\right)$ are shown in Figure \ref{fig:FinalTree}.

The foregoing discussion reveals that the \emph{k}-d tree includes ``less than" and ``greater than" children but no ``equal" children.  For this reason, duplicate $\left(x,y,z\right)$ tuples must be removed from the data prior to building the \emph{k}-d tree.  After the $k$ initial merge sorts have reordered the $xyz$-, $yzx$- and $zxy$-index arrays, each index array is traversed once in order to discard all but one index from each set of contiguous indices that reference identical $\left(x,y,z\right)$ tuples.  In order that adjacent indices reference identical $\left(x,y,z\right)$ tuples, the $k$ initial merge sorts employ $x$:$y$:$z$, $y$:$z$:$x$ and $z$:$x$:$y$ super keys instead of keys that are merely the individual $x$-, $y$- and $z$-coordinates.  If the $k$ initial merge sorts employed keys that were only the individual $x$-, $y$- and $z$-coordinates, adjacent indices within an index array could reference non-identical $\left(x,y,z\right)$ tuples for which one or more, but not all, of the $x$-, $y$- and $z$-coordinates were identical.  The $x$:$y$:$z$, $y$:$z$:$x$ and $z$:$x$:$y$ super keys guarantee that each group of identical $\left(x,y,z\right)$ tuples is indexed by a set of contiguous indices within each index array.  These super keys enable the removal of duplicate $\left(x,y,z\right)$ tuples via one pass through each index array that discards adjacent indices that reference identical $\left(x,y,z\right)$ tuples.

It is possible to optimize the use of the temporary index array such that only one temporary index array is required and such that the $xyz$-, $yzx$- and $zxy$-index arrays may be reused to avoid allocation of new index arrays at each level of recursion.  This optimization operates as follows.  The $xyz$-index array is copied to the temporary index array.  Then the $yzx$-index array is partitioned in $x$ and the result is stored in the two halves of the $xyz$-index array.  Next, the $zxy$-index array is partitioned in $x$ and the result is stored in the two halves of the $yzx$-index array.  Finally, the temporary index array is copied to the $zxy$-index array.  This optimization permutes the $xyz$-, $yzx$- and $zxy$-index arrays cyclically at each level of recursion, as is required to cycle the keys in the order $x,y,z,x,y,z...$ for successive levels of the \emph{k}-d tree.  Moreover, it guarantees that the $x$:$y$:$z$, $y$:$z$:$x$ or $z$:$x$:$y$ super key that is required for partitioning at a particular level of the \emph{k}-d tree is always specified by the median element of the $xyz$-index array.  The additional computational cost of this index array optimization is the copying of one index array at each level of recursion. (The Appendix describes a more sophisticated approach to permuting the index arrays.)

Recursive partitioning occurs for $\log_2 \left(n\right)$ levels of the nascent \emph{k}-d tree.  The computational complexity of this \emph{k}-d tree-building algorithm includes an $O\left(kn \log n\right)$ term for the $k$ initial merge sorts plus an $O\left(kn \log n\right)$ term for partitioning or copying $n$ elements of $k$ index arrays at each of the $\log_2 \left(n\right)$ levels of recursion.  This $O\left(kn \log n\right)$ \emph{k}-d tree-building algorithm requires storage for a Tuples array of $n$ \emph{k}-dimensional tuples, plus an $n$-element temporary array, plus $k$ $n$-element index arrays.  The Tuples array is immutable.  The index and temporary arrays are ephemeral and are no longer required after construction of the \emph{k}-d tree.

\subsection{Parallel Execution}
\label{sec:parallel-execution}

The merge-sorting function \cite{Sedgewick} and the $O\left(kn \log n\right)$ \emph{k}-d tree-building function both subdivide index arrays and process non-overlapping halves of each index array via recursive calls to these functions.  Hence, these functions (or Java methods) are amenable to parallel execution via multiple threads that occurs as follows.

One thread executes a recursive call of the method; this thread is designated as the parent thread.  The parent thread subdivides one or more index arrays, then calls the method recursively to process the lower and upper halves of each index array.  The parent thread does not execute the recursive call that processes the lower half of each index array; instead, it launches a child thread to execute that recursive call.  After launching the child thread, the parent thread executes the recursive call that processes the upper half of each index array, then waits for the child thread to finish execution of the recursive call that processes the lower half of each index array.

For a balanced \emph{k}-d tree, the number of threads $q$ that are required by this thread-allocation strategy is $q = 2^d \approx n / 2$ where $d$ represents the deepest level of recursion and $n$ represents the number of tuples.  A large number of tuples would require a prohibitively large number of threads; hence, child threads are launched to only the maximum level of recursion that is allowed by the number of available threads.  Beyond this maximum level of recursion, the current thread processes both halves of each index array.

Two threads permit launching a child thread at the first level of recursion.  Four threads permit launching child threads at the first two levels of recursion. Eight threads permit launching child threads at the first three levels of recursion, \emph{etc}.  Because child threads are launched at the highest levels of the tree (\emph{i.e.,} the lowest levels of recursion), each thread processes the maximum possible workload.  Because index arrays are subdivided by their median elements at each level of recursion, all threads share the workload equally.

A disadvantage of this thread allocation strategy is that it limits the number of threads to an integer power of two.  Because the level of recursion determines the number of threads, it is not possible to employ, for example, three or ten threads.  An advantage of this thread allocation strategy is that it is simple and robust because synchronization involves only a parent thread and one child thread.

\subsection{Results for the $\mathitbf{O\left(kn \; \boldsymbol{\log} \; n\right)}$ Algorithm}
\label{sec:knlogn-results}

\begin{figure}[h]
\centering
\centerline{\includegraphics*[trim = {1.42in, 3.85in, 1.37in, 1.99in}, clip, width=\columnwidth]{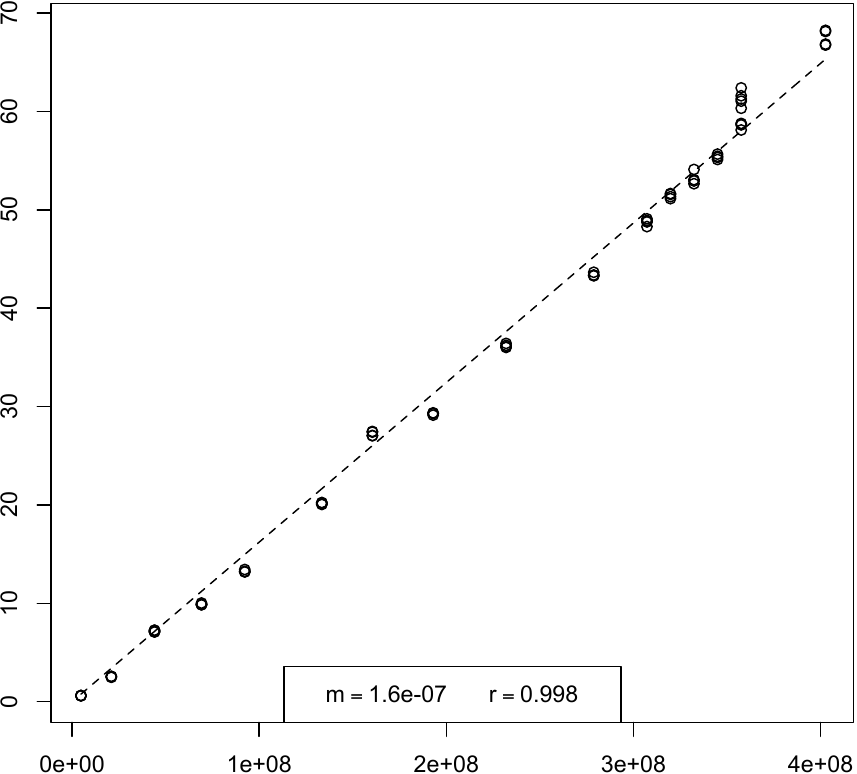}}
\caption{The total of merge sorting, duplicate tuple removal, and \emph{k}-d tree-building times (seconds) is plotted versus $\mathitbf{n \; \boldsymbol{\log_2} \; \left(n\right)}$ for the application of the $\mathitbf{O\left(kn \; \boldsymbol{\log} \; n\right)}$ \emph{k}-d tree-building algorithm to $\mathitbf{2^{18} \le n \le 2^{24}}$ $\mathitbf{\left(x,y,z,w\right)}$ tuples of randomly-generated 32-bit integers.}
\label{fig:BuildingTime}
\end{figure}

The $O\left(kn \log n\right)$ \emph{k}-d tree-building algorithm was implemented in the Java language, and the single-threaded performance of the merge sorting, duplicate tuple removal, and \emph{k}-d tree-building methods was measured using a 2.3 GHz Intel i7 processor.  Figure \ref{fig:BuildingTime} shows the total time in seconds that was required to perform the initial merge sorts, remove the duplicate tuples, and build the \emph{k}-d tree, plotted versus $n \log_2 \left(n\right)$ for $2^{18} \le n \le 2^{24}$ $\left(x,y,z,w\right)$ tuples of randomly-generated 32-bit integers. The dashed line depicts the least-squares fit of the total time $t$ to the function $t = mn \log_2 \left(n\right)$ where $m$ is the slope of the line.  The correlation coefficient $r = 0.998$ indicates an adequate least-squares fit; hence, the execution times are proportional to $n \log_2 \left(n\right)$.  The question of whether these execution times are proportional to $k$ is discussed in Section \ref{sec:discussion} of this article.

The $O\left(kn \log n\right)$ \emph{k}-d tree-building algorithm was parallelized via Java threads and its performance was measured for one to eight threads using a 2.3 GHz Intel quad-core i7 processor.  Figure \ref{fig:BuildingParallel} shows the total time in seconds that was required to perform the initial merge sorts, remove the duplicate tuples, and build the \emph{k}-d tree, plotted versus the number of threads $q$ for $n=2^{24}$ $\left(x,y,z,w\right)$ tuples of randomly-generated 32-bit integers. The dashed curve depicts the least-squares fit of the total time $t$ to the equation
\begin{equation}
t =  t_\mathrm{s} + \frac{t_1}{q} + m_\mathrm{c}\left(q - 1\right)
\label{eq:gunther}
\end{equation}
The correlation coefficient $r=0.9867$ indicates an acceptable least-squares fit.  Equation \ref{eq:gunther} and the least-squares fit are discussed in Sections \ref{sec:discussion} and \ref{sec:least-squares} respectively.

\section{Comparative Performance}
\subsection{The $\mathitbf{O\left(n \; \boldsymbol{\log} \; n\right)}$ Algorithm}
\label{sec:nlogn-algorithm}

In order to understand the performance of the $O\left(kn \log n\right)$ \emph{k}-d tree-building algorithm relative to that of other algorithms, the performance was compared to an $O\left(n \log n\right)$ \emph{k}-d tree-building algorithm that incorporates an $O\left(n\right)$ median-finding algorithm \cite{Blum}.  The application of the $O\left(n \log n\right)$ \emph{k}-d tree-building algorithm to sort $\left(x,y,z\right)$ tuples is described as follows.

\begin{figure}[h!]
\centering
\centerline{\includegraphics*[trim = {1.42in, 3.85in, 1.37in, 1.94in}, clip, width=\columnwidth]{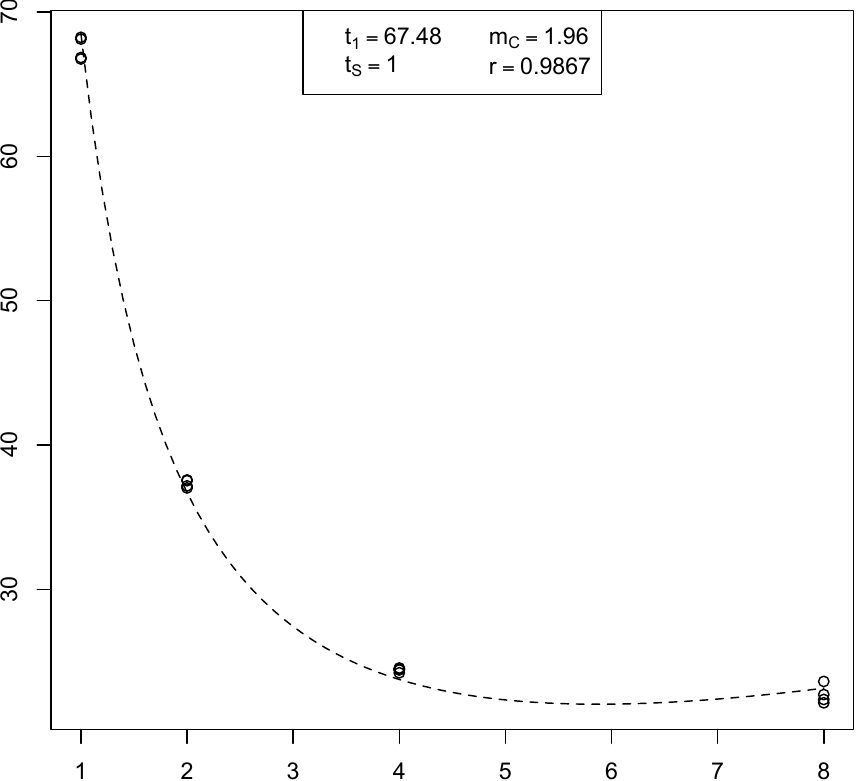}}
\caption{The total of merge sorting, duplicate tuple removal and \emph{k}-d tree-building times (seconds) is plotted versus the number of threads for the application of the $\mathitbf{O\left(kn \; \boldsymbol{\log} \; n\right)}$ \emph{k}-d tree-building algorithm to  $\mathitbf{n = 2^{24}}$ $\mathitbf{\left(x,y,z,w\right)}$ tuples of randomly-generated 32-bit integers.}
\label{fig:BuildingParallel}
\end{figure}

First, an index array is created and merge sorted in $x$, $y$ or $z$ via one of the $x$:$y$:$z$, $y$:$z$:$x$ and $z$:$x$:$y$ super keys; the choice of super key is arbitrary.  The initial merge sort does not reorder the tuples array; instead, it reorders the index array whose elements are indices into the tuples array.  Next, duplicate $\left(x,y,z\right)$ tuples are removed via one pass through the index array, as discussed in Section \ref{sec:knlogn-algorithm} of this article.

The subsequent \emph{k}-d tree-building step partitions the index array recursively.  At each level of recursion, the median element of the index array is found in $O\left(n\right)$ time using the $x$:$y$:$z$, $y$:$z$:$x$ or $z$:$x$:$y$ super key that is appropriate to that level of recursion.  A convenient feature of the $O\left(n\right)$ median-finding algorithm is that the index array is partitioned relative to the median element during the search for the median element.  Hence, once the median element has been found, a reference to the $\left(x,y,z\right)$ tuple that the median element specifies is stored in the root of the nascent \emph{k}-d tree, as shown in Figure \ref{fig:FinalTree}.  The lower and upper halves of the index array are processed recursively to create the ``less than" and ``greater than" subtrees of the root of the \emph{k}-d tree.  The $O\left(n \log n\right)$ \emph{k}-d tree-building method processes non-overlapping halves of the index array via recursive calls to this method.  Hence, this method is amenable to parallel execution via multiple threads in the manner that was explained in Section \ref{sec:parallel-execution} of this article.

Recursion terminates when the index array comprises one, two or three elements.  In the case of one element, a reference to the corresponding tuple is stored in a new node of the \emph{k}-d tree.  For two elements, a reference to the tuple that corresponds to the first element is stored in a new node of the \emph{k}-d tree, then the super keys of the two elements are compared to decide whether to reference the tuple that corresponds to the second element from the ``less than" or ``greater than" child of that node.  For three elements, the index array is sorted via insertion sort \cite{Bentley2} to determine which tuple to reference from a new node of the \emph{k}-d tree and which tuples to reference from the children of that node.

Recursive partitioning occurs for $\log_2 \left(n\right)$ levels of the nascent \emph{k}-d tree.  The computational complexity of this \emph{k}-d tree-building algorithm includes an $O\left(n \log n\right)$ term for the initial merge sort plus another $O\left(n \log n\right)$ term for partitioning $n$ elements of the index array at each of the $\log_2 \left(n\right)$ levels of recursion.  This $O\left(n \log n\right)$ \emph{k}-d tree-building algorithm requires storage for a Tuples array of $n$ \emph{k}-dimensional tuples, plus an $n$-element index array, plus an $n/2$-element temporary array.  The Tuples array is immutable.  The index and temporary arrays are ephemeral and are no longer required after construction of the \emph{k}-d tree.

\subsection{Results for the $\mathitbf{O\left(n \; \boldsymbol{\log} \; n\right)}$ Algorithm}
\label{sec:nlogn-results}

\begin{figure}[h!]
\centering
\centerline{\includegraphics*[trim = {1.42in, 3.86in, 1.37in, 2.00in}, clip, width=\columnwidth]{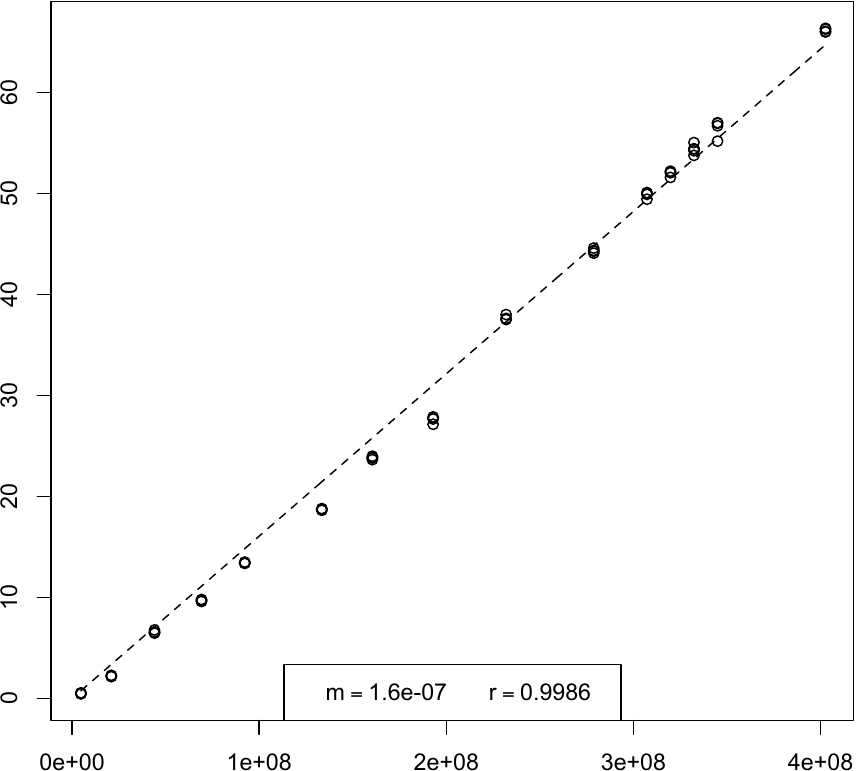}}
\caption{The total of merge sorting, duplicate tuple removal, and \emph{k}-d tree-building times (seconds) is plotted versus $\mathitbf{n \; \boldsymbol{\log_2} \; \left(n\right)}$ for the application of the $\mathitbf{O\left(n \; \boldsymbol{\log} \; n\right)}$ \emph{k}-d tree-building algorithm to $\mathitbf{2^{18} \le n \le 2^{24}}$ $\mathitbf{\left(x,y,z,w\right)}$ tuples of randomly-generated 32-bit integers.}
\label{fig:ComparativeTime}
\end{figure}

The $O\left(n \log n\right)$ \emph{k}-d tree-building algorithm was implemented in the Java language, and the single-threaded performance of the merge sorting, duplicate tuple removal, and \emph{k}-d tree-building methods was measured using a 2.3 GHz Intel i7 processor.  Figure \ref{fig:ComparativeTime} shows the total time in seconds that was required to perform the initial merge sort, remove the duplicate tuples, and build the \emph{k}-d tree, plotted versus $n \log_2 \left(n\right)$ for $2^{18} \le n \le 2^{24}$ $\left(x,y,z,w\right)$ tuples of randomly-generated 32-bit integers.  The dashed line depicts the least-squares fit of the total time $t$ to the function $t = mn \log_2 \left(n\right)$ where $m$ is the slope of the line.  The correlation coefficient $r = 0.9986$ indicates an adequate least-squares fit.

\newpage

\begin{figure}[h]
\centering
\centerline{\includegraphics*[trim = {1.42in, 3.85in, 1.37in, 2.00in}, clip, width=\columnwidth]{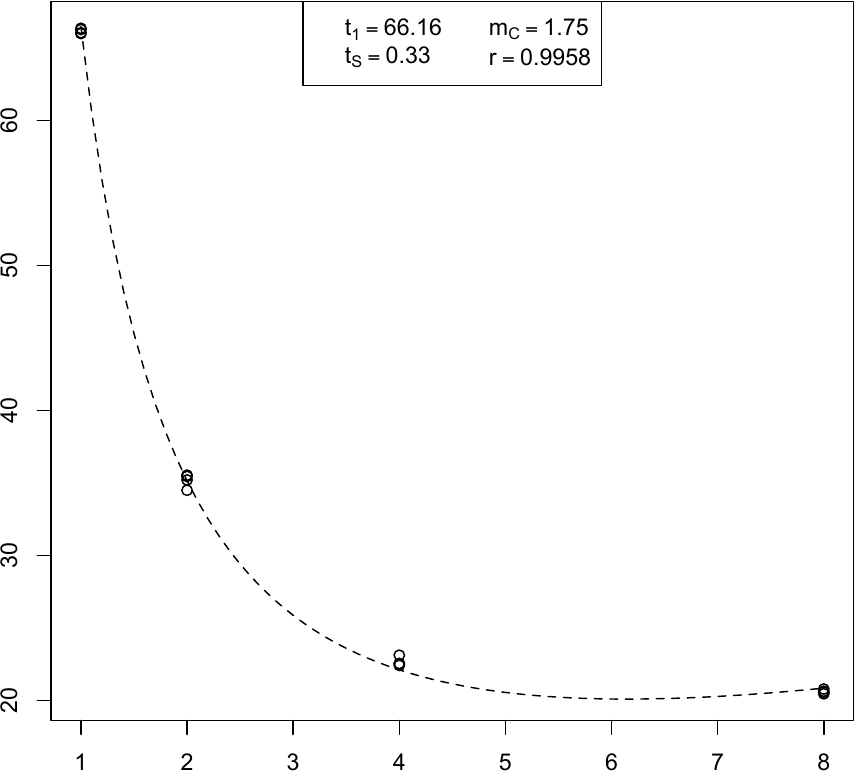}}
\caption{The total of merge sorting, duplicate tuple removal, and \emph{k}-d tree-building times (seconds) is plotted versus the number of threads for the application of the $\mathitbf{O\left(n \; \boldsymbol{\log} \; n\right)}$ \emph{k}-d tree-building algorithm to  $\mathitbf{n = 2^{24}}$ $\mathitbf{\left(x,y,z,w\right)}$ tuples of randomly-generated 32-bit integers.}
\label{fig:ComparativeParallel}
\end{figure}

The $O\left(n \log n\right)$ \emph{k}-d tree-building algorithm was parallelized via Java threads and its performance was measured for one to eight threads using a 2.3 GHz Intel quad-core i7 processor.  Figure \ref{fig:ComparativeParallel} shows the total time in seconds that was required to perform the initial merge sort, remove the duplicate tuples, and build the \emph{k}-d tree, plotted versus the number of threads $q$ for $n=2^{24}$ $\left(x,y,z,w\right)$ tuples of randomly-generated 32-bit integers. The dashed curve depicts the least-squares fit of the total time $t$ to Equation \ref{eq:gunther}.  The correlation coefficient $r=0.9958$ indicates an acceptable least-squares fit.

\section{Discussion}
\label{sec:discussion}

Figures \ref{fig:BuildingTime} and \ref{fig:ComparativeTime} demonstrate that the execution times of the $O\left(kn \log n\right)$ and $O\left(n \log n\right)$ \emph{k}-d tree-building algorithms are proportional to $n \log_2 \left(n\right)$.  Figures \ref{fig:BuildingParallel} and \ref{fig:ComparativeParallel} show that the \emph{k}-d tree-building algorithms scale for multiple threads.  For either algorithm, the execution by eight threads on an Intel quad-core i7 processor, which supports concurrent execution of two threads per core, increases the execution speed by approximately three times relative to the speed of one thread.  The execution time $t$ does not adhere to the Amdahl \cite{Amdahl} model
\begin{equation}
t =  t_\mathrm{s} + \frac{t_1}{q}
\label{eq:amdahl}
\end{equation}
but rather to the model that is expressed by Equation \ref{eq:gunther}
\begin{equation*}
t =  t_\mathrm{s} + \frac{t_1}{q} + m_\mathrm{c}\left(q - 1\right)
\end{equation*}
In this equation, $q$ is the number of threads, $t_\mathrm{s}$ represents the time required to execute the serial or non-parallelizable portion of the algorithm, $t_1$ represents the time required to execute the parallelizable portion of the algorithm via one thread, and $m_\mathrm{c}\left(q - 1\right)$ models an additional limitation to the performance of multi-threaded execution that the Amdahl model fails to capture.

This additional limitation to performance may occur due to cache contention in a shared cache wherein multiple threads attempt to access the cache simultaneously.  The cache miss term $m_\mathrm{c}\left(q - 1\right)$ of Equation \ref{eq:gunther} models the performance limitation, which results from cache contention, as a linear function of $q$ \cite{Gunther}.

Differentiating Equation \ref{eq:gunther} with respect to $q$ yields
\begin{equation}\frac{dt}{dq} = m_\mathrm{c} - \frac{t_1}{q^2}
\label{eq:derivative}
\end{equation}
Setting $dt / dq$ to zero in Equation \ref{eq:derivative} and solving for $q$ predicts that the minimum execution time should occur at $q = \sqrt{t_1 / m_\mathrm{c}}$ threads.  Substituting into this equation for $q$ the respective values of $t_1$ and $m_\mathrm{c}$ that were obtained via least-squares fitting for the $O\left(kn \log n\right)$ and $O\left(n \log n\right)$ algorithms predicts minima at $q = 5.87$ and $q = 6.15$ threads respectively.  These minima are depicted in Figures \ref{fig:BuildingParallel} and \ref{fig:ComparativeParallel} that predict decreased performance of both \emph{k}-d tree building algorithms for greater than eight threads.  The decreased performance for as few as eight threads is a consequence of the relatively large values for $m_\mathrm{c}$ (1.96 and 1.75 seconds per thread respectively) that were obtained via least-squares fitting.  Figure \ref{fig:Optimized} in Section \ref{sec:least-squares} plots execution time data obtained using an Intel 6-core i7 processor that supports concurrent execution of two threads per core. The least-squares fit to the data demonstrates a minimum execution time at $q = \sqrt{t_1 / m_\mathrm{c}} = \sqrt{5.92 / 0.0627} = 9.72$ threads.

\begin{figure}[h]
\centering
\centerline{\includegraphics*[trim = {1.42in, 3.86in, 1.37in, 2.00in}, clip, width=\columnwidth]{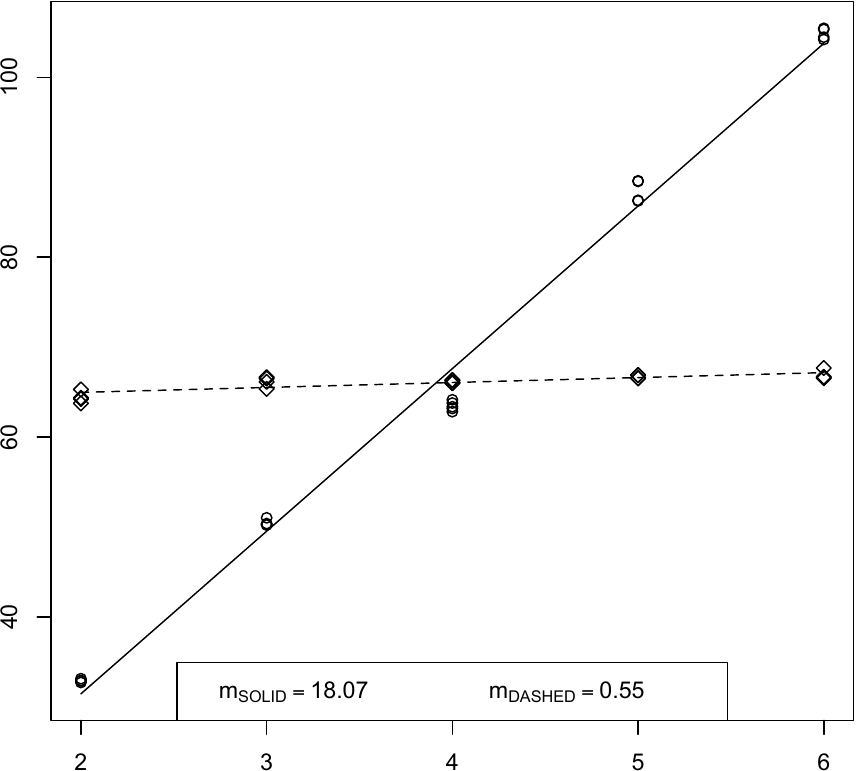}}
\caption{The total of merge sorting, duplicate tuple removal, and \emph{k}-d tree-building times (seconds) is plotted versus the number of dimensions $\mathitbf{k}$ for the application of the $\mathitbf{O\left(kn \; \boldsymbol{\log} \; n\right)}$ algorithm (solid line and circles) and the $\mathitbf{O\left(n \; \boldsymbol{\log} \; n\right)}$ algorithm (dashed line and diamonds) to build a \emph{k}-d tree from $\mathitbf{n = 2^{24}}$ \emph{k}-dimensional tuples of randomly-generated 32-bit integers.}
\label{fig:Dimensions}
\end{figure}

The similar parallelizable times $t_1$ for both algorithms (67.48 and 66.16 thread-seconds) indicate that their single-threaded performance is about equal.  This effect is due to a fortuitous choice of $k=4$ that specifies test data that comprise $\left(x,y,z,w\right)$ tuples.  Because the execution time of the $O\left(kn \log n\right)$ algorithm should be proportional to $k$ but the execution time of the $O\left(n \log n\right)$ algorithm should not, these two algorithms are expected to have unequal performance for a different choice of $k$.  In order to test this hypothesis, each algorithm was utilized to build five different \emph{k}-d trees.  For each \emph{k}-d tree, $2^{24}$ \emph{k}-dimensional tuples of randomly-generated 32-bit integers were created using a different value of $k=2, 3, 4, 5, 6$.  The performance of each algorithm was measured using a single thread of a 2.3 GHz Intel i7 processor.  The results are shown in Figure \ref{fig:Dimensions}.

Figure \ref{fig:Dimensions} shows the total time in seconds that was required to perform the initial merge sorts, remove the duplicate tuples, and build the \emph{k}-d tree via the $O\left(kn \log n\right)$ and $O\left(n \log n\right)$ algorithms for $n = 2^{24}$ \emph{k}-dimensional tuples of randomly-generated 32-bit integers, plotted versus the number of dimensions $k=2, 3, 4, 5, 6$.  This figure demonstrates that the execution time of the $O\left(kn \log n\right)$ algorithm is proportional to $k$ but the execution time of the $O\left(n \log n\right)$ algorithm is not.  In this figure, the slope of the solid line ($m_{_\mathrm{SOLID}}=18.07$ seconds per dimension) indicates that for $2^{24}$ tuples, each additional dimension increases the execution time of the $O\left(kn \log n\right)$ algorithm by 18 seconds.  For $k=4$, the two algorithms have equal performance.

In Figure \ref{fig:Dimensions}, the slope of the dashed line ($m_{_\mathrm{DASHED}}=0.55$ seconds per dimension) suggests that the execution time of the $O\left(n \log n\right)$ algorithm might be proportional to $k$.  However, this apparent proportionality is related to the time required to generate random 32-bit integers.

The storage requirements of the $O\left(kn \log n\right)$ and $O\left(n \log n\right)$ algorithms differ.  Although both algorithms require storage for a tuples array of $n$ \emph{k}-dimensional tuples, the $O\left(kn \log n\right)$ algorithm requires storage for an $n$-element temporary array plus $k$ $n$-element index arrays, whereas the $O\left(n \log n\right)$ algorithm requires storage for an $n/2$-element temporary array plus only one $n$-element index array.
  
The $O\left(n\right)$ median-finding algorithm is complicated and requires careful implementation to achieve optimum performance.  For example, the median-finding algorithm utilizes a sorting algorithm to sort large numbers of five-element arrays.  For the initial implementation of the $O\left(n\right)$ median-finding algorithm, these small arrays were sorted via the merge sort algorithm that is used for the initial sort of the index array \cite{Sedgewick}.  That merge sort algorithm is not optimized for sorting small arrays and hence resulted in poor performance of the $O\left(n\right)$ median-finding algorithm and consequent poor performance of the $O\left(n \log n\right)$ \emph{k}-d tree-building algorithm.  Replacing the merge sort algorithm with an insertion sort algorithm that is optimized for sorting small arrays \cite{Bentley2} conferred a 30 percent improvement to the performance of the $O\left(n \log n\right)$  \emph{k}-d tree-building algorithm.  Section \ref{sec:median} describes a preferred alternative to insertion sort.

\section{Conclusion}
\label{sec:conclusion}

The \emph{k}-d tree-building algorithm proposed in this article guarantees a computational complexity of $O\left(kn \log n\right)$ for $n$ points and $k$ dimensions.  For $k=4$, the performance of this algorithm equals the performance of an $O\left(n \log n\right)$ \emph{k}-d tree-building algorithm that employs an $O(n)$ median-finding algorithm.  For either algorithm, an improvement in performance by a factor of three relative to single-threaded performance is achieved via parallel execution by eight threads of an Intel quad-core i7 processor that supports concurrent execution of two threads per core.

\subsection*{Source Code}

 Source code that implemented the $O\left(n \log n\right)$ and $O\left(kn \log n\right)$ \emph{k}-d tree-building algorithms and that was provided with prior versions of this article has been superseded by the source code described in Section \ref{sec:source-code}.

\subsection*{Acknowledgements}

The author thanks Paul McJones, Gene McDaniel, Joseph Wearing and John Robinson for helpful comments.

\section{Appendix}
\label{sec:appendix}

\hfill \break \textbf{Russell A. Brown and John A. Robinson} \hfill \break

Six modifications that optimize the performance of the $O\left(n \log n\right)$ and $O\left(kn \log n\right)$ \emph{k}-d tree-building algorithms are described below.

\subsection{Optimized Super Key Comparison}

Both tree-building algorithms depend on the comparison of super keys. This comparison is performed by comparing respective portions of two super keys serially in a loop, beginning with the most significant portion of each super key, proceeding towards the least significant portion of each super key, and exiting the loop as soon as the super keys differ in their respective portions. 

A more efficient comparison may be accomplished by comparing the most significant portions of the super keys prior to entering the loop and thereby avoiding execution of the loop for the common case where the two super keys differ in their most significant portions. This simple change significantly improves the performance of both tree-building algorithms by eliminating the overhead of loop execution.

\subsection{Optimized Merge Sort}

Both tree-building algorithms depend on merge sort that copies elements between arrays \cite{Neumann}.  An optimized merge sort minimizes copying via four variants of merge sort \cite{Sedgewick}.

As discussed in Section \ref{sec:parallel-execution} of this article, merge sort uses a separate thread to process each half of the index array recursively so that the two halves of the index array are processed concurrently by two threads.  However, subsequent to this concurrent recursive processing, only one thread merges the two halves of the index array. 

Concurrent merging may be accomplished using two threads.  One thread merges beginning at the lowest address of the lower half of the index array and at the highest address of the upper half of the index array \cite{Sedgewick}; the addresses move towards one another as they are updated. A second thread merges beginning at the highest address of the lower half of the index array and at the lowest address of the upper half of the index array; the addresses move away from one another as they are updated.

Concurrent merging and minimization of copying improve the performance of both tree-building algorithms.

\newpage
\subsection{Optimized Partitioning}

As discussed in Section \ref{sec:parallel-execution} of this article, the $O\left(kn \log n\right)$ tree-building algorithm uses a separate thread to process each half of the index array recursively so that the two halves of the index array are processed concurrently by two threads.  However, prior to this concurrent recursive processing, only one thread partitions the index array into two halves. Concurrent partitioning may be performed using two threads. One thread begins at the lowest address of the the index array and ascends. A second thread begins at the highest address of the index array and descends.

As discussed in Section \ref{sec:knlogn-algorithm} of this article, the $O\left(kn \log n\right)$ algorithm partitions $k-1$ index arrays and copies one index array at each level of the tree.  An optimization avoids copying the index array and instead only partitions the $k-1$ index arrays to achieve $O\left[\left(k - 1\right)n \log n\right]$ performance. This optimization requires precomputing an array of $\log_2 n$ vectors, where each vector contains $k+1$ elements that specify permutation of the index arrays for a given level of the tree.

 Concurrent partitioning and avoidance of copying improve the performance of the $O\left(kn \log n\right)$ tree-building algorithm.

\subsection{Precomputing the Modulus}

Both tree-building algorithms select the most significant portion of the super key via a modulus during each recursive call of the tree-building function. This modulus may be precomputed and stored in an array of $\log_2 n$ elements, where each element contains the modulus for a given level of the tree, thus eliminating $n / 2 - \log_2 n$ modulus computations.

\subsection{Optimized Median Selection}
\label{sec:median}

The $O\left(n \log n\right)$ tree-building algorithm uses insertion sort \cite{Bentley2} to select a median. The performance of this tree-building algorithm may be improved by median selection via an explicit decision tree \cite{Stepanov} instead of via insertion sort that requires more comparisons and memory accesses.

\subsection{Deferred Merge Sort and Partitioning}

The $O\left(kn \log n\right)$ tree-building algorithm presorts the data in each of $k$ dimensions prior to building the tree and then partitions the data in each of $k-1$ dimensions about a median at each level of the tree. The performance of this algorithm may be improved by presorting in only \emph{one} dimension prior to building the tree.  Sorting in each additional dimension is deferred until the level of the tree where the sort is necessary. Moreover, partitioning the data in a given dimension is deferred until after that dimension has been sorted.

Deferred sorting and partitioning are illustrated by the following algorithm for building a \emph{k}-d tree from a set of $\left(x,y,z,w\right)$ tuples.  First, the tuples are presorted using their $x$:$y$:$z$:$w$ super keys to populate the $xyzw$ index array.

At the first level of the nascent tree, the $xyzw$ index array is split at its median element to create two half-index arrays. These half-index arrays are then sorted using a $y$:$z$:$w$:$x$ super key to populate two $yzwx$ half-index arrays.

At the second level of the tree, each $yzwx$ half-index array is split at its median element to create four quarter-index arrays. These quarter-index arrays are then sorted using a $z$:$w$:$x$:$y$ super key to populate four $zwxy$ quarter-index arrays.  Also at this second level of the tree, each $xyzw$ half-index array is partitioned about a $zwxy$-median to create four $xyzw$ quarter-index arrays.

At the third level of the tree, each $zwxy$ quarter-index array is split at its median element to create eight eighth-index arrays. These eighth-index arrays are then sorted using a $w$:$x$:$y$:$z$ super key to populate eight $wxyz$ eighth-index arrays.  Also at this third level of the tree, each $xyzw$ quarter-index array and each $yzwx$ quarter-index array is partitioned about a $wxyz$-median to create eight $xyzw$ eighth-index arrays and eight $yzwx$ eighth-index arrays respectively.

At this point, the data have been sorted in each of the $k$ dimensions, so beginning with the fourth level of the tree, the $O\left(kn \log n\right)$ \emph{k}-d tree-building algorithm proceeds as described in Section \ref{sec:knlogn-algorithm} of this article. The performance of this algorithm improves due to deferred sorting and partitioning to the degree discussed below.

Without deferred sorting, the computational complexity of presorting in $k$ dimensions is $O\left(kn \log_2 n\right)$. With deferred sorting, the computational complexity is
\begin{equation*}
O\left( n \log_2 n + 2 \frac{n}{2} \log_2 \frac{n}{2} + 4 \frac{n}{4} \log_2 \frac{n}{4} + 8 \frac{n}{8} \log_2 \frac{n}{8} + ... \right)
\end{equation*}
that reduces to
\begin{equation*}
O\{ n\left[ \log_2 n + \left(\log_2 n - 1\right) + \left(\log_2 n - 2\right) + \left(\log_2 n - 3\right) + ... \, \right] \}
\end{equation*}
and refactors to
\begin{equation*}
O\{ kn \log_2 n - n \left[1 + 2 + 3 + ... + \left(k - 1 \right) \right] \}
\end{equation*}
that sums to
\begin{equation*}
O\{ kn \log_2 n - n \left[ \frac{ k \left(k - 1 \right) }{2} \right] \}
\end{equation*}
and simplifies to
\begin{equation*}
O\left[ kn \left( \log_2 n - \frac{k - 1}{2} \right) \right]
\end{equation*}
Hence, deferred sorting confers to the sorting performance a fractional improvement of
\begin{equation}
\label{eq:presort4}
1 - \frac{ kn \left( \log_2 n - \frac{k-1}{2} \right) }{ kn \log_2 n } = \frac{k - 1}{2 \log_2 n}
\end{equation}

Without deferred partitioning, the computational complexity of partitioning $n$ elements of $k-1$ index arrays at each of the $\log_2 n$ levels of the tree is $O\left[\left(k - 1\right)n \log_2 n\right]$. With deferred partitioning, the computational complexity is
\begin{equation*}
O\{ n \left[ 0 + 1 + 2 + ... + \left(k - 1\right) + \left(k - 1\right) \left( \log_2 n - k \right) \right] \}
\end{equation*}
that sums and simplifies to
\begin{equation*}
O\left[ \left(k - 1\right) n \left( \log_2 n - \frac{k}{2} \right) \right]
\end{equation*}
Hence, deferred partitioning confers to the partitioning performance a fractional improvement of
\begin{equation}
\label{eq:partition3}
1 - \frac{ \left(k - 1\right) n \left( \log_2 n - \frac{k}{2} \right) }{ \left(k - 1\right) n \log_2 n } = \frac{k}{2 \log_2 n}
\end{equation}

Equations \ref{eq:presort4} and \ref{eq:partition3} predict that deferral improves the performance of sorting and partitioning by a factor that is inversely proportional to $\log_2 n$ and directly proportional to $k - 1$ and $k$ respectively.

\subsection{Performance Measurements}

The performance of optimized $O\left(n \log n\right)$ and $O\left(kn \log n\right)$ \emph{k}-d tree-building algorithms that benefit from the six above-described modifications was compared to the performance of the non-optimized $O\left(kn \log n\right)$ algorithm described in Section \ref{sec:knlogn-algorithm} of this article. C++ implementations were benchmarked using an Intel quad-core i7 processor executing eight threads to process $2^{24}$ 4-dimensional randomly-generated tuples of 64-bit integers.  The optimized $O\left(n \log n\right)$ and $O\left(kn \log n\right)$ algorithms achieved 28 percent and 26 percent performance improvements respectively, relative to the non-optimized $O\left(kn \log n\right)$ algorithm.  For this benchmark test case, equations \ref{eq:presort4} and \ref{eq:partition3} predict performance improvements of 6 percent and 8 percent respectively.  Hence, the optimizations discussed above (other than deferred sorting and partitioning) significantly improve the performance of the $O\left(kn \log n\right)$ algorithm and presumably the performance of the $O\left(n \log n\right)$ algorithm as well.

A second benchmark used an Intel quad-core i7 processor executing eight threads to process $2^{24}$ 8-dimensional randomly-generated tuples of 64-bit integers.  The optimized $O\left(n \log n\right)$ and $O\left(kn \log n\right)$ algorithms achieved 65 percent and 34 percent performance improvements respectively, relative to the non-optimized $O\left(kn \log n\right)$ algorithm.   The greater performance improvement of the $O\left(n \log n\right)$ algorithm relative to the $O\left(kn \log n\right)$ algorithm is due to the fact that the computational complexity of the $O\left(kn \log n\right)$ algorithm is proportional to the number of dimensions $k$ whereas the computational complexity of the $O\left(n \log n\right)$ algorithm is independent of $k$, as demonstrated by Figure \ref{fig:Dimensions}.

\subsection{Nearest Neighbors}
The \emph{k}-d tree may be used to find the nearest neighbor to a query point in $O\left(\log n\right)$ expected time \cite{Friedman}. The nearest-neighbor search algorithm finds the $m$ nearest neighbors to a query point and sorts those $m$ nearest neighbors according to their distances to the query point via a priority queue \cite{Sedgewick}. In addition, the nearest neighbors can be used to find the \emph{reverse} nearest neighbors to a query point, where the reverse nearest neighbors to a given point are defined as the set of points to which that point is a nearest neighbor \cite{Korn}.

The first step of the reverse-nearest-neighbors search algorithm finds the $m$ nearest neighbors to each point in the \emph{k}-d tree in $O\left(n \log n\right)$ expected time by walking the tree recursively and treating each point in the tree as a query point. This nearest-neighbors search algorithm executes in parallel via multiple threads in the manner explained in Section \ref{sec:parallel-execution} for the \emph{k}-d tree-building algorithms. As the $m$ nearest neighbors to each query point are found, the second step of the reverse-nearest-neighbors search algorithm appends that query point to a unique reverse-nearest-neighbors list for each of the $m$ nearest neighbors. The reverse-nearest-neighbors search algorithm executes in parallel via multiple threads and requires a unique lock for each reverse-nearest-neighbors list because any point in the \emph{k}-d tree may be a reverse nearest neighbor to any other point in the \emph{k}-d tree.

\begin{figure}[h]
\centering
\centerline{\includegraphics*[trim = {1.42in, 3.86in, 1.37in, 2.00in}, clip, width=\columnwidth]{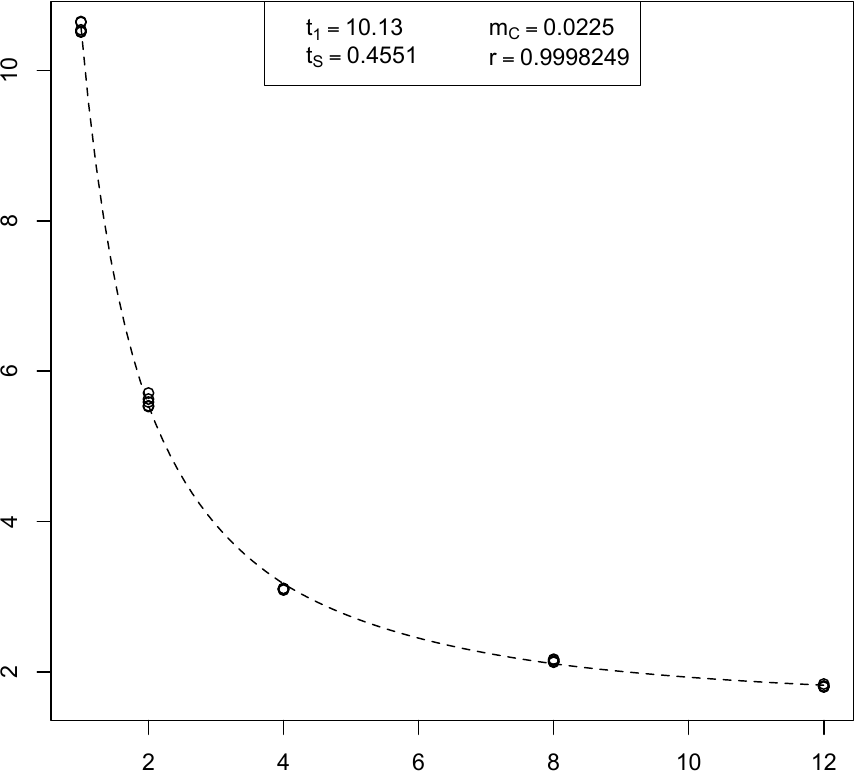}}
\caption{The total execution time (seconds) is plotted versus the number of threads for the application of the nearest-neighbor and reverse-nearest-neighbor search algorithms to a \emph{k}-d tree built from $\mathitbf{n = 2^{22}}$ $\mathitbf{\left(x,y,z \right)}$ tuples of randomly-generated 64-bit integers.}
\label{fig:Neighbors}
\end{figure}

The execution times of the nearest-neighbors and reverse-nearest-neighbors search algorithms were measured for one to 12 threads using a 3.2 GHz Intel six-core i7 processor.  Figure \ref{fig:Neighbors} shows the time in seconds required to search the \emph{k}-d tree for nearest neighbors and reverse nearest neighbors, plotted versus the number of threads $q$ for $n=2^{22}$ $\left(x,y,z \right)$ tuples of randomly-generated 64-bit integers.

\subsection{Source Code for Optimized Algorithms}
\label{sec:source-code}

Provided under the BSD 3-Clause License are C++ implementations of the optimized $O\left(n \log n\right)$ and $O\left(kn \log n\right)$ algorithms described as follows.

The ``test\textbf{\_}kdtree.cpp" and ``test\textbf{\_}kdmap.cpp" programs build and test a \emph{k}-d tree and a \emph{k}-d tree-based key-to-multiple-value map respectively. The tree and map may be built via either the $O\left(n \log n\right)$ or the $O\left(kn \log n\right)$ algorithm.

The ``test\textbf{\_}kdmap.cpp" program may achieve higher performance by storing a tuple in each node of the \emph{k}-d tree, instead of in the ``Tuples" array depicted in Figure \ref{fig:IndexArrays}, to avoid one degree of indirection when accessing the tuple. In addition, this program may be compiled to build a \emph{k}-d tree instead of a \emph{k}-d tree-based key-to-multiple-value map.

The ``test\textbf{\_}kdtree.cpp" program tests the ``knTreeKnlogn.h", ``kdTreeNlogn.h", ``kdTreeNode.h", ``kdTreeHeapSort.h", and ``kdTreeMergeSort.h" files.

The ``test\textbf{\_}kdmap.cpp" program tests the ``knMapKnlogn.h", ``kdMapNlogn.h", ``kdMapNode.h", ``kdMapHeapSort.h", and ``kdMapMergeSort.h" files.

The ``kdTreeNode.h" and ``kdMapNode.h" files implement algorithms that search a \emph{k}-d tree and map respectively (1) for all points that lie inside a \emph{k}-dimensional hyper-rectangular region \cite{Bentley}, (2) for the nearest neighbors to a query point \cite{Friedman}, and (3) for the nearest neighbors and reverse nearest neighbors to all points in the \emph{k}-d tree \cite{Korn}.

The algorithms that build a \emph{k}-d tree and search a hyper-rectangular region use multiple threads. The algorithm that searches for the nearest neighbors to a query point uses a single thread. The algorithm that searches for the nearest neighbors and reverse nearest neighbors to all points in the \emph{k}-d tree uses multiple threads.

\newpage

\subsection{Least-Squares Curve Fitting}
\label{sec:least-squares}

Figures \ref{fig:BuildingParallel} and \ref{fig:ComparativeParallel} plot the least-squares fits of Java execution time data for the non-optimized $O\left(kn \log n\right)$ and $O\left(n \log n\right)$ \emph{k}-d tree-building algorithms respectively to the model defined by Equation \ref{eq:gunther} in Section \ref{sec:knlogn-results}
\begin{equation*}
t =  t_\mathrm{s} + \frac{t_1}{q} + m_\mathrm{c}\left(q - 1\right)
\end{equation*}
where $t$ is the execution time, $q$ is the number of threads, $t_\mathrm{s}$ is the non-parallelizable execution time, $t_1$ is the parallelizable execution time for a single thread, and $m_\mathrm{c}$ extends the Amdahl model to include the effects of cache contention.

To facilitate least-squares curve fitting, the above equation is rewritten as
\begin{equation}
\label{eq:leastsquares}
t =  t_\mathrm{s} + t_1 f + m_\mathrm{c} g
\end{equation}
where $ f = 1 / q $ and $ g = q - 1 $.

A least-squares fit \cite{Thomas} of the $\left( f_\mathrm{i},  g_\mathrm{i}, t_\mathrm{i} \right)$ tuples of measured data is performed by constructing the error function
\begin{equation}
\label{eq:error}
E = \sum_{i=1}^{n} \left ( t_i -  t_\mathrm{s} - t_1 f_i - m_\mathrm{c} g_i \right )^2
\end{equation}
and then obtaining the partial derivatives of the error function with respect to $ t_\mathrm{s} $, $ t_1 $ and $ m_\mathrm{c} $ as follows.
\begin{equation}
\label{eq:partials}
\begin{matrix}
\\ \partial E / \partial t_\mathrm{s} = -2 \sum \left ( t_i -  t_\mathrm{s} - t_1 f_i - m_\mathrm{c} g_i \right ) = 0
\\
\\ \partial E / \partial t_1 = -2 \sum \left ( t_i -  t_\mathrm{s} - t_1 f_i - m_\mathrm{c} g_i \right ) f_i = 0
\\
\\ \partial E / \partial m_\mathrm{c} = -2 \sum \left ( t_i -  t_\mathrm{s} - t_1 f_i - m_\mathrm{c} g_i \right ) g_i = 0
\end{matrix}
\end{equation}

Each partial derivative equals zero at the minimum of the error function. Rearranging Equation \ref{eq:partials} yields the following three simultaneous linear equations that may be solved for the three unknowns $ t_\mathrm{s} $, $ t_1 $ and $ m_\mathrm{c} $ via Cramer's rule.
\begin{equation}
\label{eq:normal}
\begin{matrix}
\\  n t_\mathrm{s}+ t_1 \sum f_i + m_\mathrm{c} \sum g_i = \sum t_i
\\
\\ t_\mathrm{s} \sum f_i + t_1 \sum f_{i}^2 + m_\mathrm{c} \sum f_i g_i = \sum t_i f_i
\\
\\ t_\mathrm{s} \sum g_i + t_1 \sum f_i g_i + m_\mathrm{c} \sum g_{i}^2 = \sum t_i g_i
\end{matrix}\end{equation}

The correlation coefficient $r_\mathrm{fgt}$ for the least-squares fit is a coefficient of linear multiple correlation that is calculated from the linear correlation coefficients $r_\mathrm{ft}$, $r_\mathrm{gt}$, and $r_\mathrm{fg}$ as follows \cite{Spiegel}.
\begin{equation}
\label{eq:correlation}
\begin{matrix}
\\ r_\mathrm{ft} = \frac{ n \sum {f_i t_i} - \sum f_i \sum t_i } { \sqrt { \left [ n \sum {f_i}^2 - \left ( \sum f_i \right )^2 \right ] \left [ n \sum {t_i}^2 - \left ( \sum t_i \right )^2 \right ] } }
\\
\\ r_\mathrm{gt} = \frac{ n \sum {g_i t_i} - \sum g_i \sum t_i } { \sqrt { \left [ n \sum {g_i}^2 - \left ( \sum g_i \right )^2 \right ] \left [ n \sum {t_i}^2 - \left ( \sum t_i \right )^2 \right ] } }
\\
\\ r_\mathrm{fg} = \frac{ n \sum {f_i g_i} - \sum f_i \sum g_i } { \sqrt { \left [ n \sum {f_i}^2 - \left ( \sum f_i \right )^2 \right ] \left [ n \sum {g_i}^2 - \left ( \sum g_i \right )^2 \right ] } }
\\
\\ r_\mathrm{fgt} = \frac { r_\mathrm{ft}^2 + r_\mathrm{gt}^2 - 2 r_\mathrm{ft} r_\mathrm{gt} r_\mathrm{fg} } { 1 - r_\mathrm{fg}^2  }
\end{matrix}
\end{equation}

For a number of threads $q > 2$, Equations \ref{eq:normal} and \ref{eq:correlation} use the \emph{average} values $ \hat{f}_{i}^{q} $ and $ \hat{g}_{i}^{q} $ respectively in place of $ f_i $ and $ g_i $ because the number of threads $q$ is not constant across all levels of the nascent \emph{k}-d tree, as illustrated by the following example.

A \emph{k}-d tree built from $2^{22} $ tuples requires 21 levels of recursion. Building the tree via a single thread executes one thread at all 21 levels of recursion.  Building the tree via two threads executes two threads at all 21 levels of recursion. Building the tree via four threads executes two threads at the first level and four threads at the remaining 20 levels of recursion. Building the tree via eight threads executes two threads at the first level, four threads at the second level, and eight threads at the remaining 19 levels of recursion. Building the tree via 12 threads executes two threads at the first level, four threads at the second level, eight threads at the third level, and 12 threads at the remaining 18 levels of recursion. For building the tree via 12 threads, 16 threads are created and hence are potentially executable at the remaining 18 levels of recursion but the Intel 6-core i7 processor actually executes at most 12 threads concurrently.

Given the execution by different numbers of threads at different levels of recursion, the average values $ \hat{f}_{i}^{q} $ and $ \hat{g}_{i}^{q} $ are calculated using $q=4$, $q=8$ and $q=12$ as follows.

For $q = 4$ the respective average values $ \hat{f}_{i}^{4} $ and $ \hat{g}_{i}^{4}$ of $ f_i $ and $ g_i $ are calculated as
\begin{equation}
\begin{matrix}
\\ \hat{f}_{i}^{4} = \frac{1/2 + 20/4} {21} = 0.2619
\\
\\ \hat{g}_{i}^{4} =  \frac{2 + 20 \times{4}} {21} - 1 = 2.90
\end{matrix}
\label{eq:average4}
\end{equation}

For $q = 8$ the respective average values $ \hat{f}_{i}^{8} $ and $ \hat{g}_{i}^{8}$ of $ f_i $ and $ g_i $ are calculated as
\begin{equation}
\begin{matrix}
\\ \hat{f}_{i}^{8} = \frac{1/2 + 1/4 + 19/8} {21} = 0.1488
\\
\\ \hat{g}_{i}^{8} =  \frac{2 + 4 + 19 \times{8}} {21} - 1 = 6.52
\end{matrix}
\label{eq:average8}
\end{equation}

For $q = 12$ the respective average values $ \hat{f}_{i}^{12} $ and $ \hat{g}_{i}^{12}$ of $ f_i $ and $ g_i $ are calculated as
\begin{equation}
\begin{matrix}
\\ \hat{f}_{i}^{12} = \frac{1/2 + 1/4 + 1/8 + 18/12} {21} = 0.1131
\\
\\ \hat{g}_{i}^{12} =  \frac{2 + 4 + 8 + 18 \times{12}} {21} - 1 = 9.95
\end{matrix}
\label{eq:average8}
\end{equation}

\begin{figure}[h]
\centering
\centerline{\includegraphics*[trim = {1.42in, 3.86in, 1.37in, 2.00in}, clip, width=\columnwidth]{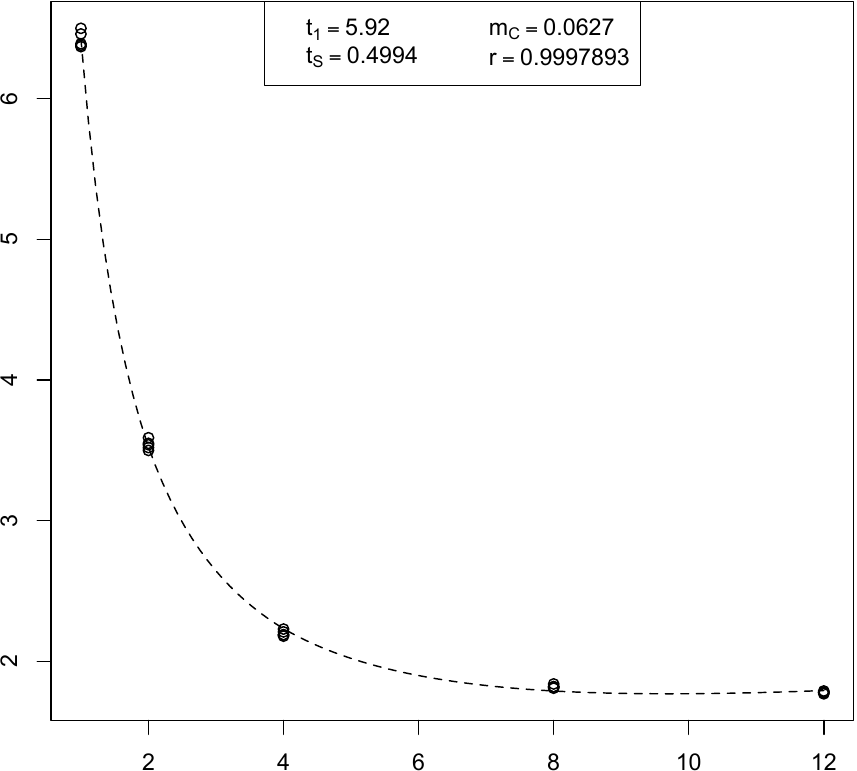}}
\caption{The total of merge sorting, duplicate tuple removal, and \emph{k}-d tree-building times (seconds) is plotted versus the number of threads for the application of the optimized $\mathitbf{O\left(n \; \boldsymbol{\log} \; n\right)}$ \emph{k}-d tree-building algorithm to  $\mathitbf{n = 2^{22}}$ $\mathitbf{\left(x,y,z \right)}$ tuples of randomly-generated 64-bit integers.}
\label{fig:Optimized}
\end{figure}

The execution time of the optimized $O\left(n \log n\right)$ \emph{k}-d tree-building algorithm was measured for one to 12 threads using a 3.2 GHz Intel 6-core i7 processor.  Figure \ref{fig:Optimized} shows the total time in seconds that was required to perform the initial merge sort, remove the duplicate tuples, and build the \emph{k}-d tree, plotted versus the number of threads $q$ for $n=2^{22}$ $\left(x,y,z \right)$ tuples of randomly-generated 64-bit integers.

The dashed curve of Figure \ref{fig:Optimized} plots the equation
\begin{equation}
\label{eq:plot}
t =  t_\mathrm{s} + t_1 \hat{f}_{u:v} + m_\mathrm{c} \hat{g}_{u:v}
\end{equation}
where the \emph{average} values $\hat{f}_{u:v}$ and $\hat{g}_{u:v}$ are calculated using $q$ in the interval $u < q \le v$ as follows.

For $2 < q \leq 4$, $\hat{f}_{2:4}$ and $\hat{g}_{2:4}$ are calculated as
\begin{equation}
\begin{matrix}
\\ \hat{f}_{2:4} = \frac{1/2 + 20 / q} {21}
\\
\\ \hat{g}_{2:4} =  \frac{2 + 20q} {21} - 1
\end{matrix}
\label{eq:plot4}
\end{equation}

For $4 < q \leq 8$, $\hat{f}_{4:8}$ and $\hat{g}_{4:8}$ are calculated as
\begin{equation}
\begin{matrix}
\\ \hat{f}_{4:8} = \frac{1/2 + 1/4 + 19 / q} {21}
\\
\\ \hat{g}_{4:8} =  \frac{2 + 4 + 19q} {21} - 1
\end{matrix}
\label{eq:plot8}
\end{equation}

For $8 < q \leq 12$, $\hat{f}_{8:12}$ and $\hat{g}_{8:12}$ are calculated as
\begin{equation}
\begin{matrix}
\\ \hat{f}_{8:12} = \frac{1/2 + 1/4 + 1/8 + 18 / q} {21}
\\
\\ \hat{g}_{8:12} =  \frac{2 + 4 + 8 + 18q} {21} - 1
\end{matrix}
\label{eq:plot12}
\end{equation}

\bibliographystyle{abbrv}
\bibliography{building_balanced_kd_tree.bib}

\end{document}